\documentclass[12pt,a4paper]{article}
\usepackage{amsmath,amssymb,amsthm}
\usepackage[margin=1.0in]{geometry}
\usepackage{cite}
\usepackage{graphicx}
\usepackage{enumerate}
\allowdisplaybreaks
\usepackage[colorlinks=true
,urlcolor=blue
,anchorcolor=blue
,citecolor=blue
,filecolor=blue
,linkcolor=blue
,menucolor=blue
,pagecolor=blue
,linktocpage=true
,pdfproducer=medialab
,pdfa=true
]{hyperref}
\numberwithin{equation}{section}

\def\g{\gamma}
\def\d{\delta}
\def\e{\epsilon}

\def\k{\kappa}
\def\l{\lambda}
\def\m{\mu}
\def\n{\nu}

\def\r{\rho}

\def\s{\sigma}

\def\ph{\phi}

\def\Th{\Theta}

\def\S{\Sigma}

\def\L{\Lambda}

\def\be{\begin{equation}}
\def\ee{\end{equation}}
\def\bea{\begin{eqnarray}}
\def\eea{\end{eqnarray}}

\def\pa{\partial}

\def\lp{\left(}
\def\rp{\right)}
\def\ls{\left[}
\def\rs{\right]}
\def\nn{\nonumber}
\def\ie{{\it i.e., }}

\makeatletter
\renewcommand\section{\@startsection {section}{1}{\z@}%
	{-3.5ex \@plus -1ex \@minus -.2ex}
	{2.3ex \@plus.2ex}%
	{\normalfont\large\bfseries}}
\renewcommand\subsection{\@startsection{subsection}{2}{\z@}%
	{-3.25ex\@plus -1ex \@minus -.2ex}%
	{1.5ex \@plus .2ex}%
	{\normalfont\bfseries}}
\makeatother

\begin{document}

\begin{center}
\addtolength{\baselineskip}{.5mm}
\thispagestyle{empty}
\begin{flushright}
\end{flushright}

\vspace{20mm}

{\Large \bf Charged rotating black strings \\ in  Einsteinian cubic gravity}
\\[15mm]
{Hamid R. Bakhtiarizadeh\footnote{h.bakhtiarizadeh@kgut.ac.ir}}
\\[5mm]
{\it Department of Nanotechnology, Graduate University of Advanced Technology,\\ Kerman, Iran}

\vspace{20mm}

{\bf  Abstract}
\end{center}

We construct, for the first time, charged rotating black string solutions in four-dimensional Einsteinian cubic gravity, which are asymptotically anti–de Sitter. By assuming that the solutions are completely regular at the horizon and studying their near-horizon behavior, we find some thermodynamic properties, which can be accessed analytically. We compute independently the Hawking temperature, the Wald entropy, the mass, the angular momentum, the charge, and the electrostatic potential of the solutions, analytically. Using these, we show that the first law of thermodynamics for rotating black strings is exactly satisfied in both charged and uncharged cases. We also observe that, in the absence of Maxwell field, some of the solutions have positive specific heat, which makes them thermodynamically stable. 

\vfill
\newpage


\section{Introduction}\label{int}

One of the most relevant approaches in modification of general relativity (GR) is to take into account the higher-curvature terms. In addition to the Lovelock densities \cite{Lovelock:1971yv}, there exists another class of theories, so-called quasitopological gravities (QTG) \cite{Myers:2010ru,Oliva:2010eb}, which admit analytic static black holes in dimensions larger than 4. The most general theory of gravity up to a cubic order in curvature is called generalized quasitopological gravity \cite{Hennigar:2017ego,Bueno:2017sui,Ahmed:2017jod,Bueno:2017qce,Bueno:2019ycr}, whose static and spherically symmetric vacuum solutions are fully described by a single function which comes from second-order linearized equations around maximally symmetric backgrounds \cite{Ahmed:2017jod,Bueno:2019ycr,Bueno:2019ltp,Hennigar:2017umz,Mir:2019rik,Mir:2019ecg}. In this theory, the Lovelock and quasitopological gravities, which are determined by an algebraic equation, have been recovered as special cases in four dimensions.

Recently, it was shown that up to cubic order in curvature, there is a theory so-called Einsteinian cubic gravity (ECG) as the most general higher-curvature modification of Einstein gravity, which only propagates the usual transverse and traceless graviton on maximally symmetric backgrounds in general dimensions \cite{Bueno:2016xff}. This theory admits nonhairy single-function generalizations of the Schwarzschild black hole similar to the Lovelock and quasi-topological theories in dimensions larger than 4 \cite{Bueno:2016lrh,Hennigar:2016gkm}. 

Here we review some attempts in finding the solutions in the context of ECG theory. In \cite{Frassino:2020zuv}, the authors show that static charged black holes in four-dimensional ECG differ from their counterparts in GR in several aspects: (i) nonuniqueness of regular solutions, (ii) noncompliance with the extremality bound, and (iii) the absence of an inner horizon. In \cite{Hennigar:2018hza}, using a continued fraction ansatz, an analytic approximation for a spherically symmetric black hole solution to ECG has been found. ECG theory has been regarded as a holographic toy model of a nonsupersymmetric conformal field theory (CFT) in three dimensions in \cite{Bueno:2018xqc}. Also, the first generalizations of the Einstein gravity Taub-NUT/bolt solutions for any higher-curvature theory in four dimensions have been represented in \cite{Bueno:2018uoy}. In \cite{Burger:2019wkq}, using on-shell amplitudes, a rotating black hole solution in ECG theory is derived. Recently, slowly rotating black hole solutions of ECG in four dimensions with flat and anti-de Sitter (AdS) asymptotes were constructed and various physical properties of the solutions were studied \cite{Adair:2020vso}. In \cite{Cano:2019ozf}, new solutions of Einsteinian cubic gravity coupled to a Maxwell field that describe the near-horizon geometry of charged rotating black holes are obtained.

In the present work, we will study asymptotically AdS charged rotating black strings \cite{Lemos:1994xp,Lemos:1995cm} in the context of ECG theory. It has been shown that for charged rotating toroidal black hole solutions only configurations that are supersymmetric are: (i) the nonrotating electrically charged naked singularities, and (ii) an extreme rotating toroidal black hole with zero electric and magnetic charges \cite{Lemos:2000wp}. The solutions of charged rotating black strings, which are indeed cylindrically symmetric black holes, have also been studied in \cite{Cai:1996eg,Cardoso:2001vs} and are generalized to include higher dimensions as black strings/branes in \cite{Awad:2002cz}. The thermodynamic properties of solutions are also investigated in \cite{Dehghani:2002rr,Dehghani:2002jh}. Rotating black string solutions have been also explored in the presence of nonlinear electrodynamics \cite{Hendi:2010kv,Hendi:2013mka}, dilaton gravity \cite{Dehghani:2004sa,Sheykhi:2008rk}, holographic superfluids \cite{Brihaye:2010ce}, $ f(R) $ gravity \cite{Sheykhi:2013yga}, and mimetic gravity \cite{Sheykhi:2020fqf,Bakhtiarizadeh:2021pyo}.

The structure of the paper is organized as follows. In the next section, we will study, in detail, uncharged solutions both asymptotically and near a black string horizon. We then obtain, using a Taylor expansion around the horizon, exact expressions for the mass and surface gravity. We also compute the angular momentum of solutions, as well as the Wald entropy, and show that the first law of black string thermodynamics holds exactly. In Sec. \ref{ch}, we investigate the charged solutions. As the uncharged case, by studying the near-horizon behavior of solutions, we get the mass, angular momentum, as well as some thermodynamic properties. We also find the electrostatic potential and total charge of solutions. Having had these quantities, we easily check the first law of black string thermodynamics in the presence of Maxwell field and find an exact agreement. Section \ref{con} is also devoted to summary and concluding remarks.

\section{Asymptotically AdS uncharged solutions}\label{unch}

The action of four-dimensional ECG theory in the presence of cosmological constant reads,
\be
S=\frac{1}{16\pi G}\int d^4x \sqrt{-g}(R-2\L-2 G^2 \l {\cal P}),\label{action}
\ee
where $ G $ is the Newton gravitational constant, $ R $ represents the Ricci scalar, $ \L=-3/{l^2} $ is the negative cosmological constant of AdS space,\footnote{Note that the asymptotically de–Sitter solutions can be obtained by simply taking $ l\rightarrow il $ \cite{Awad:2002cz}.} and the cubic-in-curvature correction to the Einstein-Hilbert action is incorporated in
\bea
{\cal P}=12 R_{a}{}^{c}{}_{b}{}^{d} R_{c}{}^{e}{}_{d}{}^{f} R_{e}{}^{a}{}_{f}{}^{b}+R_{ab}{}^{cd} R_{cd}{}^{ef} R_{ef}{}^{ab}-12 R_{abcd}
R^{ac} R^{bd}+8 R_{a}{}^{b} R_{b}{}^{c} R_{c}{}^{a}.\label{ECGterms}
\eea
We will assume the dimensionless ECG coupling constant to be non-negative throughout the paper, \ie $ \l \geq 0 $. Einstein gravity is also recovered by setting $ \l = 0 $.

Our aim here is to construct the asymptotically AdS charged rotating black string solutions of ECG theory and investigate their properties. Therefore, we assume the metric of four-dimensional spacetime with cylindrical or toroidal horizons can be written as \cite{Lemos:1994xp,Lemos:1995cm,Lemos:2000wp} 
\bea
ds^2=-f(r)g^2(r)\lp \Xi dt -a d\phi \rp ^2+\frac{1}{f(r)}dr^2+\frac{r^2}{l^4} \lp a dt -\Xi l^2 d\phi \rp ^2+\frac{r^2}{l^2}dz^2,\label{met}
\eea
where
\be
\Xi=\sqrt{1+a^2/l^2}.\label{sileq}
\ee
The constants $ a $ and $ l $ have dimensions of length and can be interpreted as the rotation parameter and the AdS radius, respectively. In the following, we are going to study the solutions of ECG with cylindrical symmetry. This implies that the spacetime admits a commutative two-dimensional Lie group $ G_2 $. The ranges of the time and radial coordinates are $ -\infty<t<\infty, 0\leq r<\infty $, and the topology of the horizon can be regarded as follows: \begin{enumerate}[(i)] \item the flat torus $ T^2 $ with topology $ S^1\times S^1 $ (\ie $ G_2=U(1) \times U(1) $) and the ranges $ 0\leq \phi<2\pi,0\leq z<2\pi l $, which describes a closed black string, \item the cylinder with topology $ \mathbb{R}\times S^1 $ (\ie $ G_2=\mathbb{R} \times U(1) $) and the ranges $ 0\leq \phi<2\pi,-\infty<z<\infty $, which describes a stationary black string, \item the infinite plane with topology $ \mathbb{R}^2 $ and the ranges $ -\infty<\phi<\infty,-\infty<z<\infty $, which does not rotate.\end{enumerate}
We consider the topology (ii) throughout the paper.\footnote{For the case of a toroidal horizon, the entropy, mass, angular momentum, and charge of the string are obtained from their respective densities by multiplying them by $ 2\pi l $ \cite{Dehghani:2002rr}.} Now, we are in a position to evaluate the field equations of action (\ref{action}) on the ansatz (\ref{met}) and find the corresponding equations for the functions $ g(r) $ and $ f(r) $. Here, we will use a method introduced in \cite{Bueno:2016lrh,Hennigar:2016gkm} for static and spherically symmetric spacetimes. By considering the action as a functional of these functions, $ S[g,f] $, one finds
\be
\frac{\d S[g,f]}{\d g}=\frac{\d S[g,f]}{\d f}=0\Leftrightarrow {\cal E}_{tt}={\cal E}_{rr}={\cal E}_{t\phi}=0.
\ee
Here $ {\cal E}_{tt} $, $ {\cal E}_{rr} $, and $ {\cal E}_{t\phi} $ are, respectively, the $ tt $, $ rr $, and $ t\phi $ components of the corresponding field equations. Notice that the metric (\ref{met}) is obtained after a boost along the $ (t,\phi) $ coordinates from the static black brane solution. The boost is only locally defined, so the global structure of the metric is different, but this explains why the metric takes this particular form and why it is not necessary to solve other components of the equations of motion. This shows that the equations for $ f(r) $ and $ g(r) $ can be obtained from the action functional $ S[g,f] $ without the need to compute the full non-linear equations explicitly. It can be seen that the action $ S[g,f] $ can be written as
\bea
S[g,f]=\frac{1}{8\pi G}\int dr g(r).\left\{-\frac{1}{3}\L r^3-r f-G^2 \l \ls4 f'^3-12 f f' f''-24\frac{f^2 \lp f'-r f''\rp}{r^2}\rs\right\}'
\eea
plus some terms involving at least two derivatives of $ g $. Here, a prime denotes a derivative with respect to $ r $. By variation of the above action with respect to $ g $ and $ f $, one can get the equations of them. It can be seen from that, $ g $ is multiplied by a total derivative. Therefore, $ \d_{g}S=0 $ can be solved by setting
\be
g'(r)=0. 
\ee
As a result, the ECG theory admits solutions characterized by a single function $ f(r) $. So we set $ g = 1 $.\footnote{Without loss of generality we can set $ g(r) = 1 $, for simplicity. In general, one can choose $ g = 1/\sqrt{f_{\infty}} $, where $ f_{\infty}=\lim_{r\rightarrow\infty} f(r) $, to normalize the speed of light on the boundary or in the dual CFT to be $ c = 1 $. However, we can set $ g = 1 $ by reparametrization of time ($ t $) and angular ($ \phi $) coordinates of the metric, if desired.} The equation $ \d_{g}S = 0 $ yields, after integrating once, the following equation for $ f $:
\be
-\frac{1}{3}\L r^3-r f-G^2 \l \ls4 f'^3-12 f f' f''-24\frac{f^2 \lp f'-r f''\rp}{r^2}\rs=r_0,\label{uncheom}
\ee
where $ r_0 $ is an integration constant which is related to the mass of string as $ r_0=G M $.

\subsection{Asymptotic solution}\label{unchasymp}

As can be seen from Eq. (\ref{uncheom}), when $ \l=0 $ we obtain
\be
f(r)=\frac{r^2}{l^2}-\frac{GM}{r},
\ee
where we have set $ \L=-3/{l^2} $ and $ r_0=GM $. This is, of course, nothing but the usual AdS uncharged solution of rotating black strings \cite{Lemos:1995cm,Lemos:1994xp}. When $ \l $ is turned on, the asymptotic quantities get corrected in this case. To obtain these corrections, we first examine the large-$ r $ behavior of the solution. To do so, we assume that the metric function $ f $ can be expressed as a particular solution in the form of a $ 1/r $ expansion, plus the general solution of the corresponding homogeneous equation,
\be
f=f_{1/r}+f_{h}\quad \text{with} \quad f_{1/r}(r)=\frac{r^2}{l_{\rm eff}^2}+\sum_{n=1}^{\infty}\frac{b_{n}}{r^n}.\label{asyexp}
\ee
Substituting this series expansion into Eq. (\ref{uncheom}), one finds the large-$ r $ expansion reads
\be
f_{1/r}(r)=\frac{r^2}{l_{\rm eff}^2}-\frac{G_{\rm eff} M}{r}+{\cal O}\lp r^{-3}\rp,\label{f1r}
\ee
where the effective radius of the AdS space, $ {l_{\rm eff}} $, is a solution of the equation
\be
\frac{16 G^2 \l}{l_{\rm eff}^6}-\frac{1}{l_{\rm eff}^2}+\frac{1}{l^2}=0,\label{leffeq}
\ee
and the effective gravitational constant is given by
\be
G_{\rm eff}=\frac{G}{1-48\frac{G^2 \l}{l_{\rm eff}^4}}.\label{geffeq}
\ee
On the other hand, the linearized homogeneous equation satisfied by $ f_{h}(r) $ at the large-$ r $ limit, reads
\be
f_{h}''(r)-\frac{4}{r}f_{h}'(r)-\g^2 r f_{h}(r)=0,\label{uncheq}
\ee
where
\be
\g^2=\frac{l_{\rm eff}^2}{36 G G_{\rm eff}^2 M \l}.\label{gvalue}
\ee
Here, we have kept only the leading terms in the large-$ r $ limit. The solution of Eq. (\ref{uncheq}), in the case of $ \g^2>0 $, is
\be
f_{h}^{(+)}(r)= r^{5/2}\ls A I_{5/3}\lp \frac{2\g r^{3/2}}{3}\rp+B K_{5/3}\lp \frac{2\g r^{3/2}}{3}\rp \rs,
\ee
where $ I $ and $ K $ are the modified Bessel functions of the first and second kinds, and $ A $ and $ B $ are some constants. In the limit of large $ r $ we can approximate the solution by
\be
f_{h}^{(+)}(r)\sim r^{5/2}\ls A \exp\lp \frac{2\g r^{3/2}}{3}\rp+B \exp\lp- \frac{2\g r^{3/2}}{3}\rp \rs,\label{apsol}
\ee 
and so we must set $ A = 0 $ to ensure the AdS boundary conditions are satisfied. As a result no ghost excitations can propagate to infinity. We shall see shortly that the contribution of the second term can be dismissed. If $ \g^2 < 0 $, then the homogeneous solution asymptotically takes the following form: 
\be
f_{h}^{(-)}(r)= r^{5/2}\ls C J_{5/3}\lp \frac{2\left|\g\right| r^{3/2}}{3}\rp+D Y_{5/3}\lp \frac{2\left|\g\right| r^{3/2}}{3}\rp \rs,
\ee
where $ J $ and $ Y $ are the Bessel functions of the first and second kind, and $ C $ and $ D $ are arbitrary constants. In this situation, the solution oscillates rapidly, and its amplitude becomes larger than $ r^2/l_{\rm eff}^2 $ at large $ r $. It therefore does not approach AdS asymptotically, and so we must set $ C = D = 0 $ to get rid of this homogeneous part of the solution. For the rest of our considerations, to avoid any oscillating behavior near infinity, we restrict the solutions to the constraint $ \g^2 > 0 $. Finally, we note that the particular solution (\ref{f1r}) polynomially decreases with $ 1/r $ and is the dominant part of the total solution $ f(r) =f_{1/r} + f_h^{(+)} $ for sufficiently large $ r $; we therefore neglect the term $ f_h^{(+)} $ in Eq. (\ref{apsol}) in the sequel.

\subsection{Near-horizon solution}\label{unchnear}

Horizon is defined by surface $ r = r_h $, at which $ f(r_h) = 0 $ and $ f'(r_h) \geq 0 $. By Taylor expanding around the horizon (assuming $ f $ to be completely regular there), $ f(r)=\Sigma_{n=1}^{\infty} a_n (r-r_h)^n$, and solving (\ref{uncheom}) order by order in powers of $ (r-r_h) $, the coefficients $ a_n $ can be determined. Here, $ a_n=f^{(n)}(r_h)/n! $. The general definition of surface gravity is
\be
\k=\sqrt{-\frac{1}{2}(\nabla^a \chi^b)(\nabla_a \chi_b)}.
\ee
Here, the null generator of the black string horizon is given by $ \chi=\pa_t+\Omega\pa_\phi $, where the angular velocity of the event horizon is given by
\bea 
\Omega=\frac{a}{l^2\Xi}. 
\eea
A simple calculation for the spacetime (\ref{met}) with $ g(r)=1 $ gives $ \k_g =f'(r_h)/2\Xi $. The two lowest-order equations form an algebraic system that is used to fix the mass $ M $ and surface gravity $ \k_g $ in terms of horizon radius $ r_h $, are given by
\be
\frac{r_h^3}{l^2}-G M-32 G^2 \l \k_g^3 \Xi^3=0,
\ee
\be
\frac{3r_h^2}{l^2}-2 \k_g \Xi r_h=0.
\ee 
Solving the above equations, we get the following quantities for surface gravity and mass of black strings
\be
\k_g=\frac{3 r_h}{2 l^2 \Xi},\label{surgra}
\ee
\be
M=\frac{r_h^3}{G l^2}\lp 1- \frac{108 G^2 \l}{l^4} \rp.\label{mass}
\ee
Note that, for higher-order equations, we can treat $ a_2 $ as a free parameter and find the $ n $th coefficient $ a_{n} $, in terms of $ a_2 $, and get a family of solutions with only one free parameter $ a_2 $. For the value of $ a_2 $, we are able to construct numerically the solution up to a sufficiently large $ r $, for which the solution approaches to the asymptotic expansion in (\ref{asyexp}) \cite{Bueno:2016lrh}.

\subsection{Thermodynamics}\label{unchthermo}

We begin this section with a discussion of the black string entropy. In a higher curvature theory of gravity, the Bekenstein-Hawking entropy is modified by additional terms which can be obtained using the Wald entropy formula \cite{Wald:1993nt,Iyer:1994ys},
\be
{\sf S}=-2\pi \int_{H} d^2x\sqrt{\g}\frac{\d {\cal L}}{\d R_{abcd}}\e_{ab}\e_{cd},\label{wald}
\ee 
where $ \frac{\d {\cal L}}{\d R_{abcd}} $ is the Euler-Lagrange derivative of gravitational Lagrangian, $ \g $ is the determinant of the induced metric on the horizon, and $ \e_{ab} $ is the binormal of the horizon, normalized to satisfy $ \e_{ab} \e^{ab}=-2 $, which can be derived from the following relation: 
\be
\triangledown_a \chi_b=\k_g\e_{ab}.\label{binormal}
\ee
Using (\ref{binormal}), one can easily find the nonzero components for the antisymmetric binormal of the horizon, that are 
\bea
\e_{tr}=-\e_{rt}=-\Xi,\nn\\\e_{r\phi}=-\e_{\phi r}=-a.
\eea  
The result for entropy now reads \cite{Bueno:2016lrh},
\bea
{\sf S}&=&\frac{1}{4 G} \int_{H} d^2x\sqrt{\g}\ls 1+G^2\l\lp36R_{b}{}^{e}{}_{d}{}^{f}R_{aecf}+3R_{ab}{}^{ef}R_{cdef}\right.\right.\nn\\&&\qquad\qquad\qquad\left.\left.-12R_{ac}R_{db}-24R^{ef}R_{eafc}g_{bd}+24g_{bd}R_{ce}R^{e}{}_{a} \rp\e^{ab}\e^{cd}\rs.\label{ECGentropy}
\eea
For the metric (\ref{met}) and with a cylindrical horizon placed at $ r = r_h $, one finds the following value for entropy per unit length,
\be
{\cal S}=\frac{\pi r_h^2 \Xi}{2Gl}\lp 1-\frac{108 G^2 \l}{l^4} \rp,\label{unchen}
\ee
where $ f'(r_h) =2\Xi \k_g $, and Eq. (\ref{surgra}) has been taken into account to write the final result in terms of horizon radius $ r_h $. The Hawking temperature of our solution can easily be written in terms of the horizon radius as  
\bea
T=\frac{\k_g}{2\pi}=\frac{3r_h}{4 \pi l^2\Xi}.\label{temp}
\eea
We finish this section by calculating the mass and angular momentum of black strings by using the counterterm method inspired by AdS/CFT conjecture \cite{Maldacena:1997re}. To do so, we add the Gibbons–Hawking boundary term, which removes the divergences of the action (\ref{action}). Here, the suitable boundary action is given by \cite{Bueno:2018xqc}
\be
S_b=\frac{1+48\frac{G^2 \l}{l_{\rm eff}^4}}{8\pi G}\int_{\pa \cal{M}} d^3x \sqrt{-\g} \Th,
\ee
where $ \g $ is the determinant of the induced metric on the boundary and $ \Th $ is the trace of the extrinsic curvature $ \Th_{ab} $ of the boundary. Note that $ l_{\rm eff} $ is a scale length factor that depends on $ l $ and $ \l $ that must reduce to $ l $ as $ \l $ goes to zero. That is indeed the root of Eq. (\ref{leffeq}). We use the counterterm method \cite{Henningson:1998gx,Balasubramanian:1999re} to eliminate the divergences of action. In this approach, we add some local surface integrals to the action to make it finite. This method also has been applied to the case of third-order Lovelock gravity with flat boundary $ {\cal R}_{abcd}(\g) = 0 $ \cite{Dehghani:2006dh}. The counterterms for the case of ECG, which makes the action finite up to four dimensions, are given in \cite{Bueno:2018xqc}; they are
\be
S_{ct}=\frac{1+48\frac{G^2 \l}{l_{\rm eff}^4}}{8\pi G}\int_{\pa \cal{M}} d^3x \sqrt{-\g} \lp \frac{2}{l_{\rm eff}}-\frac{l_{\rm eff}}{2} {\cal R}\rp,\label{5ctaction}
\ee
where $ {\cal R} $ is the Ricci scalar for the boundary metric $ \g $. Notice that this counterterm has exactly the same form as the Einstein gravity with zero curvature boundary, in which $ G $ and $ l $ are replaced by $ G/\lp 1+48\frac{G^2 \l}{l_{\rm eff}^4}\rp $ and $ l_{\rm eff} $, respectively. The total action can be written as a linear combination of the action of bulk, boundary, and the counterterm,
\be
S_{total}=S+S_{b}+S_{ct}.
\ee 
Having had the total finite action, one can use the Brown-York definition of stress energy-momentum tensor \cite{Brown:1992br}, by varying the action with respect to boundary metric $ \g_{ab} $, and find a divergence-free stress tensor as
\be
T^{ab}=\frac{1}{8\pi G_{\rm eff}}\lp \Th^{ab}-\Th \g^{ab}+\frac{2}{l_{\rm eff}}\g^{ab}-\frac{l_{\rm eff}}{2} {\cal G}^{ab}\rp,\label{5enmomtens}
\ee
where $ {\cal G}_{ab}={\cal R}_{ab}-{\cal R}\g_{ab}/2 $ is the Einstein tensor of boundary metric $ \g_{ab} $. For asymptotically AdS solutions with flat horizons, the only nonvanishing counterterm, is the first term in (\ref{5ctaction}), which yields the stress-energy tensor (\ref{5enmomtens}) up to the third term. Using the above stress tensor, one can define the quasilocal conserved quantities for an asymptotically AdS spacetime. The conserved charge associated to a Killing vector $ \xi_a $ is given by
\be
Q_\xi=\int_{\S} d^3x\sqrt{\s} u^a T_{ab}\xi^b,\label{consch}
\ee
where $ u_a =-N\d^{0}_{a} $, while $ N $ and $ \s $ are the lapse function and the metric of spacelike surface $ \S $, which appear in the ADM–like decomposition of the boundary metric,
\be
\g_{ij}dx^i dx^j=-N^2 dt^2+\s_{ab}\lp dx^a + V^a dt\rp \lp dx^b + V^b dt \rp.
\ee
Here also $ V^a $ is the shift vector. To obtain the total mass (energy), we should set $ \xi=\pa_t $, \ie the Killing vector conjugate to time coordinate $ t $, and to obtain the angular momentum, we should set $ \xi=\pa_{\ph} $, \ie the Killing vector conjugate to angular coordinate $ \ph $. Using the definition (\ref{consch}) for conserved charges, we find the total mass per unit length of horizon to be given by
\be
{\cal M}=\frac{1}{8l}\lp 3 \Xi^2-1 \rp M,\label{mas}
\ee
while the angular momentum per unit length of horizon is given by
\be
{\cal J}=\frac{3}{8l} \Xi a M=\frac{3}{8} \Xi\sqrt{\Xi^2-1} M,\label{ang}
\ee
where in writing the last equality, the Eq. (\ref{sileq}) has been used. Substituting the mass (\ref{mass}) into the above equations, one arrives at   
\bea
{\cal M}=\frac{r_h^3}{8G l^3}\lp 3 \Xi^2-1 \rp\lp 1- \frac{108G^2 \l}{l^4} \rp,\label{unchmass}
\eea
\bea
{\cal J}=\frac{3r_h^3}{8G l^2}\Xi\sqrt{\Xi^2-1}\lp 1- \frac{108G^2 \l}{l^4} \rp.\label{unchang}
\eea
Using the above results, one can easily approve that the first law of thermodynamics for uncharged rotating black strings,
\be
d{\cal M}=Td{\cal S}+\Omega d{\cal J},
\ee
also holds exactly in the context of Einsteinian cubic gravity.

\subsection{Thermal stability}

The local stability, in an arbitrary ensemble, requires that the entropy $ {\cal S} $ be a convex function of their extensive variables, or its Legendre transformation must be a concave function of their intensive variables. Thus, the local stability can be explored by finding the determinant of the Hessian matrix of $ {\cal S} $ with respect to its extensive variables $ X_i $, $ {\bf H}_{X_i X_j}^{\cal S} =\pa^2 {\cal S}/\pa X_i \pa X_j $, or the determinant of the Hessian of the Gibbs function, with respect to its intensive variables $ Y_i $, $ {\bf H}_{Y_i Y_j}^{G}=\pa^2 {G}/\pa Y_i \pa Y_j $, where the thermodynamic variables of the system depend on the ensemble, which is used \cite{Cvetic:1999ne}. In the other words, the more $ X_i $ or $ Y_i $ we regard as variable parameters, the smaller is the region of stability. In the case of uncharged solutions, the entropy is a function of mass and angular momentum per unit length. In canonical ensemble, the angular momentum is the fixed parameter, and therefore, the positivity of specific heat, 
\be
C=T\lp\frac{\pa {\cal S}}{\pa T}\rp_{\cal J},
\ee
guarantees the local stability \cite{Cvetic:1999ne}. Writing the entropy (\ref{unchen}) in terms of temperature using Eq. (\ref{temp}) and then substituting it into the above expression, one finds that the specific heat parametrized by horizon radius $ r_h $ is given by
\be
C=\frac{\pi r_h^2 \Xi\lp 1-\frac{108 G^2 \l}{l^4} \rp \lp 1+\Xi^2 \rp}{2 G l\lp 5\Xi^2-4 \rp}.
\ee
It can be seen that for $ \l \leq l^4/108 G^2 $, the solutions are locally stable in the canonical ensemble.

In the grand canonical ensemble, the thermodynamic variables are the temperature and the angular velocity. So, it is more convenient to work with thermodynamic potential $ G(T, \Omega)={\cal M}-T{\cal S}-\Omega{\cal J} $, which is indeed the Legendre transformation of the energy $ {\cal M}({\cal S}, {\cal J}) $ with respect to $ {\cal S}, {\cal J} $. Here, the determinant of Hessian matrix is given by  
\be 
\vert {\bf H}_{T,\Omega}^{G}\vert=\ls \frac{\pi \Xi^2 r_h^2}{G}\lp 1-\frac{108G^2 \l}{l^4} \rp \rs^2\lp 1- \frac{\Xi^2}{2} \rp.
\ee
It is a matter of calculation to show that the solutions are locally stable in grand canonical ensemble if $ \Xi \leq \sqrt{2} $ or using Eq. (\ref{sileq}), $ a \leq l $.

\section{Asymptotically AdS charged  solutions}\label{ch}

In this section we add a Maxwell field to the action as
\be
S=\int d^4x \sqrt{-g}\ls\frac{1}{16\pi G}\lp R-2\L-2 G^2 \l {\cal P}\rp-\frac{1}{4}F_{ab}F^{ab}\rs,\label{chaction}
\ee
where $ F_{ab}=2\pa_{[a} A_{b]} $. The vector potential and nonzero components of electromagnetic field strength tensor are given by \cite{Lemos:1995cm}
\bea
&&A_{a}=A_0(r)\lp \Xi \d_{a}^{t}- a \d_{a}^{\phi} \rp;\nn\\&&F_{tr}=-F_{rt}=-\Xi A'_0(r),F_{\phi r}=-F_{r\phi}=a A'_0(r).\label{nonvanF}
\eea
Having had the electromagnetic field strength tensor, it is not difficult to show that both $ t $ and $ \phi $ components of the Maxwell equation, $ \triangledown_a F^{ab}=0 $, lead to the same equation,
\be
r A''_0 g-r A'_0 g'+2 A'_0 g =0.
\ee
Here also, as the uncharged case, we find that $ g'=0 $ and therefore set $ g=1 $. This yields the following value for the gauge potential
\be
A_0(r)=\frac{q}{4 \pi r},
\ee
where $ q $ is an integration constant, which is indeed the electric charge of the black string. As before, the equation for $ f $ can be obtained from the variation of (\ref{chaction}) with respect to $ g $ and then setting $ g=1 $, that is
\be
-\frac{1}{3}\L r^2+\frac{G Q^2}{r}-r f-G^2 \l \ls4 f'^3-12 f f' f''-24\frac{f^2 \lp f'-r f''\rp}{r^2}\rs=r_0,\label{cheom}
\ee
where we have defined $ Q^2\equiv q^2/4\pi $, and $ r_0 $ is an integration constant, which is related to the mass of the black hole as the uncharged case, $ r_0=GM $. In the next section, we will perform an asymptotic expansion around $ r\rightarrow \infty $. 

\subsection{Asymptotic solution}\label{chasymp}

In the case of charged solutions, by setting $ \l=0 $ in Eq. (\ref{uncheom}), one finds
\be
f(r)=\frac{r^2}{l^2}-\frac{GM}{r}+\frac{G Q^2}{r^2},
\ee
which is the usual asymptotically AdS charged solution of rotating black strings \cite{Lemos:1995cm,Lemos:1994xp}. As the uncharged case, one can examine the large-$ r $ behavior of the solution. To this end, we assume that $ f $ can be expressed as a particular solution in the form of a $ 1/r $ expansion, plus the general solution of the corresponding homogeneous equation. Just like the uncharged case, this expansion is given by Eq. (\ref{asyexp}). Plugging this series expansion into Eq. (\ref{cheom}), one observes that the large-$ r $ expansion takes the following form:
\be
f_{1/r}(r)=\frac{r^2}{l_{\rm eff}^2}-\frac{G_{\rm eff} M}{r}+\frac{G_{\rm eff}Q^2}{r^2}+{\cal O}(r^{-3}),\label{largerexp}
\ee
where the effective radius of the AdS space $ l_{\rm eff} $, and the effective gravitational constant $ G_{\rm eff} $ are defined by Eqs. (\ref{leffeq}) and (\ref{geffeq}), respectively. The linearized homogeneous equation satisfied by $ f_{h}(r) $ is given by
\be
f_{h}''(r)+\frac{12 M}{4 Q^2 -3 M r}f_{h}'(r)+\frac{3 M \g^2 r^2}{4 Q^2 -3 M r}f_{h}(r) =0,
\ee
at the large-$ r $ limit. Notice that when $ Q=0 $, we recover Eq. (\ref{uncheq}). Unfortunately, this equation can not be solved analytically. But due to the existence of $ \lp 4 Q^2 -3 M r \rp $ in the denominator, it is possible to expand the above equation at large $ r $ one more time and find a solution and discuss the boundary conditions. By doing so, we arrive at 
\be
f_{h}''(r)- \frac{\lp 4 Q^2+3 M r\rp \g^2}{3 M} f_{h}(r)=0.
\ee
Here, $ \g^2 $ is given by Eq. (\ref{gvalue}). The solution reads,
\be
f_{h}(r)= A' \text{Ai}\left(\frac{4 Q^2+3 M r}{3 M}\left|\gamma^2\right|^{1/3}\right)+B' \text{Bi}\left(\frac{4 Q^2+3 M r}{3 M}\left|\gamma^2\right|^{1/3}\right),
\ee
where $ \text{Ai} $ and $ \text{Bi} $ are the Airy functions of the first and second kinds, and $ A' $ and $ B' $ are some constants. In the case of $ \g^2>0 $, and at large $ r $,
\bea
f_{h}^{(+)}(r)&\sim & \left(\frac{4 Q^2+3 M r}{3 M}\gamma ^{2/3}\right)^{1/4}\times\\&&\left\{ A' \exp\ls-\frac{2}{3}\left(\frac{4 Q^2+3 M r}{3 M}\gamma ^{2/3}\right)^{3/2}\rs+B' \exp\ls \frac{2}{3}\left(\frac{4 Q^2+3 M r}{3 M}\gamma ^{2/3}\right)^{3/2}\rs \right\},\nn\label{apsol1}
\eea
and so we must set $ B' = 0 $ to ensure the AdS boundary conditions are satisfied. As a result, no ghost excitations can propagate to infinity. If $ \g^2 < 0 $, then the homogeneous solution asymptotically takes the following form at large $ r $:
\bea
f_{h}^{(-)}(r)&\sim & \left(\frac{4 Q^2+3 M r}{3 M}\gamma ^{2/3}\right)^{-1/4}\times\\&&\left\{ A' \sin\ls\frac{2}{3}\left(\frac{4 Q^2+3 M r}{3 M}\gamma ^{2/3}\right)^{3/2}\rs+B' \cos\ls \frac{2}{3}\left(\frac{4 Q^2+3 M r}{3 M}\gamma ^{2/3}\right)^{3/2}\rs \right\}.\nn
\eea
Here also the solution oscillates rapidly and therefore does not approach AdS asymptotically, and so we set $ A' = B' = 0 $. As the uncharged case, to avoid any oscillating behavior at infinity, we restrict the solutions to the constraint $ \g^2 > 0 $. Observe that all the leading asymptotic corrections to the metric come from the solution (\ref{largerexp}), while the contributions from the homogeneous equation are extremely subleading. Hence, the term proportional to $ A' $ above can be discarded from the asymptotic expansion.

\subsection{Near-horizon solution}\label{chnear}

In a similar way with the uncharged case, by making a Taylor expansion around the horizon, one finds the following equations at two lowest-order of expansion:  
\be
-G M-32 G^2 \l \k_g^3 \Xi^3+\frac{r_h^3}{l^2}+\frac{G Q^2}{r_h}=0,
\ee
\be
\frac{3r_h^2}{l^2}-2 \k_g \Xi  r_h-\frac{G Q^2}{r_h^2}=0,
\ee
which solving them yields the following values for the surface gravity, as well as mass as a function of horizon radius and charge of black strings, 
\be
\k_g=\frac{1}{2 \Xi  r_h^3}\lp\frac{3r_h^4}{l^2} -G Q^2\rp\label{chsurgra},
\ee
\be
M=\frac{r_h^3}{Gl^2}\ls1+\frac{Gl^2 Q^2}{r_h^4}-\frac{4G^2 l^2 \l}{r_h^{12}} \lp  \frac{3 r_h^4}{l^2}-G Q^2\rp^3\rs\label{chmass}.
\ee
Using these expressions, we are able to check the first law of black string thermodynamics in the presence of the Maxwell field in the next section. 
  
\subsection{Thermodynamics}\label{chthermo}

Using the entropy (\ref{ECGentropy}) and $ f'(r_h) =2\Xi \k_g $, where $ \k_g $ is given by (\ref{chsurgra}), we find the following quantity for the Wald entropy per unit length for charged solutions:
\be
{\cal S}=\frac{\pi r_h^2 \Xi}{2 G l}\ls 1- \frac{12 G^2 \l}{r_h^8} \lp \frac{3 r_h^4}{l^2}-G Q^2\rp^2\rs.
\ee
The Hawking temperature can easily be obtained in terms of charge and the horizon radius by multiplying the surface gravity (\ref{chsurgra}) with a factor $ 1/2\pi $, 
\be
T=\frac{1}{4 \pi r_h^3 \Xi }\lp\frac{3 r_h^4}{l^2} -G Q^2\rp\label{chtemp}.
\ee
The total mass and angular momentum per unit length for charged solutions can also be found by inserting Eq. (\ref{chmass}) into the Eqs. (\ref{mas}) and (\ref{ang}),
\be
{\cal M}=\frac{r_h^3}{8Gl^3}\lp 3 \Xi^2-1 \rp\ls1+\frac{Gl^2 Q^2}{r_h^4}-\frac{4G^2 l^2 \l}{r_h^{12}} \lp  \frac{3r_h^4}{l^2}-G Q^2\rp^3\rs,
\ee
\be
{\cal J}=\frac{3r_h^3}{8Gl^2} \Xi\sqrt{\Xi^2-1} \ls 1+\frac{Gl^2 Q^2}{r_h^4}-\frac{4 G^2 l^2 \l}{r_h^{12}} \lp \frac{3r_h^4}{l^2}-G Q^2\rp^3\rs.
\ee
In the following we are going to find the electric potential and total charge of black strings. The electric potential $ \Phi $, measured at infinity with respect to the horizon, is defined by \cite{Dehghani:2002rr}
\be
\Phi=A_a \chi^a\lvert_{r\rightarrow\infty}-A_a\chi^a\lvert_{r=r_h},
\ee
where, as already mentioned, $ \chi $ is the null generator of the horizon. Calculating the above expression yields the following value for electric potential
\bea
\Phi=\frac{q}{4 \pi  \Xi  r_h}=\frac{Q}{2 \sqrt{\pi } \Xi  r_h}. 
\eea
Finally, we calculate the electric charge of the black string. To this end, we first determine the electric field by considering the projections of the electromagnetic field tensor on a special hypersurface. The normal vectors to such a hypersurface are
\be
u^0 = \frac{1}{N} ,u^r = 0, u^i=-\frac{V^i}{N},
\ee
where $ N $ and $ V^i $ are the lapse function and shift vector, respectively. The electric field is also given by $ E^\m = g^{\m\r}F_{\r\n}u^{\n} $. The electric charge per unit length of black strings can be found by calculating the flux of the electric field at infinity, yielding
\bea
{\cal Q}=\frac{q \Xi}{2l}=\frac{\sqrt{\pi } Q \Xi}{l}.
\eea
Using the above expressions, one can easily show that the first law of thermodynamics for charged rotating black strings, 
\be
d{\cal M}=Td{\cal S}+\Omega d{\cal J}+\Phi d{\cal Q},
\ee
is fully satisfied in the context of ECG theory. 

\section{Conclusions}\label{con}

In this paper we have constructed generalizations of the charged rotating black strings in four-dimensional ECG with AdS asymptotic and studied some of their properties. 

We have shown that the theory admits solutions with a single function $ f(r) $, which is given by a nonlinear second-order differential equation and studied some of their thermodynamic properties which can be obtained analytically. The new solutions represent the first nontrivial four-dimensional generalizations of the asymptotically AdS charged rotating black strings in higher-order gravity, whose thermodynamic properties can be computed exactly. Using those results, we have been able to check analytically that the solutions exactly satisfy the first law of thermodynamics for black strings in both charged and uncharged cases. 

Let us close with some possible future explorations. It would be interesting to perform this kind of study, in higher dimensions. Finding the solutions in the presence of nonlinear Born-Infeld electrodynamics is also an interesting feature. Finally, a study of the solutions including Einsteinian quartic gravity would be perhaps worth pursuing.

\appendix

\section*{Acknowledgements}\addcontentsline{toc}{section}{Acknowledgements}

We would like to thank the authors of Ref. \cite{Nutma:2013zea} for developing the excellent \emph{Mathematica} package ``xTras,'' which we have used extensively for symbolic calculations.


\providecommand{\href}[2]{#2}\begingroup\raggedright
\endgroup
\end{document}